# Planetary system around Vega.

*Nick Gorkavyi, Tanya Taidakova (Schafer Corp., gorkavyi@schaferdc.com)*

We applied our earlier numerical method (Taidakova & Gorkavyi, 1999) to compute the thermal emission of a dusty resonant structure around Vega (Gorkavyi et al. 2000a, 2000b). Both the Poynting-Robertson and stellar wind drags tend to induce cometary dust inflow toward the star. The modeling has shown that a planet produces an asymmetric resonant dust belt via resonances and gravitational scattering. This feature can serve as an indicator of an invisible outer planet. Our simulations have three main steps:
1. simulation of distribution of cometary population;
2. calculation of distribution of cometary dust (see Figures 1-2);
3. determination of thermal emission of dust (see Figures 3-4).

We compared the new evidence for a ring arc at 95 AU near Vega (Koerner, Sargent & Ostroff, 2001) with our modeling. Our high resolution (2-3 AU) simulations indicate that Vega may have a outermost planet at a distance of 90-100 AU with coordinates near 18 35'16.25" (right ascension, B1950) and 38 44'15" (declination). Another symmetrical position, 18 35'14.75" and 38 44'32.5", of the planet is also possible. We can decide between these two positions after measuring the revolution direction of the resonant pattern, which must have an angular velocity of 0.6 deg/yr.

Radii of clumps (and planet's orbital radius) may be smaller: 60-75 AU (see Wilner, Holman, Ho and Kuchner, 2002). In this case the planet stay approximately on the line between the two possible points marked above and the angular velocities will be up to 1.2 deg/year. Preliminary considerations give the mass of outer planet less than ~2 Jupiter masses. Our simulations in the one planet approximation show that it must have bright emission from inner symmetrical dust disk with radius <30-40 AU. Hence, from observations by Koerner, Sargent & Ostroff, 2001 of absence of such dust disk emission, we can conclude that there is no disk out to very small distances. From our simulations we can predict very massive inner (<50-60 AU) planet(s), which destroyed inner circumstellar dust disk by gravitational scattering. If weak dusty inner clumps (see observation by Koerner, Sargent & Ostroff, 2001) have resonant link to inner planets, they must revolve around Vega much faster than the outermost arcs.

## Conclusions.

From our modeling we can make following conclusions:
1. Dust clumps near Vega are made of resonant particles (2:1; 3:1 etc);
they rotate around Vega exactly with the planet's orbital speed (precession is very small): 0.6-1.2 deg/year depending on orbital radius of parent planet (see point 2).
2. Orbital radius of the planet is approximately the same as the orbital radius of clumps (90-100 AU from observation by Koerner, Sargent & Ostroff, 2001 or 60-75 AU from observation by Wilner et al, 2002).
3. The most dense clump is ahead of the planet in the direction of the orbital motion.

4. The planet stays in the empty area farthest from the clumps.
5. This planet is the outermost massive planet in Vega's planetary system (like Neptune in Solar system).
6. The orbit of the planet may be circular or with small eccentricity.
7. Beyond the planet there exists an outer cometary belt (like TNO' belt).
8. Transparency of the inner part of Vega's planetary system is a signature of inner planet(s) with orbital radii <50-60 AU.
9. Mass of the outermost Vega's planet is near or less than ~2 Jupiter masses.

**References.**

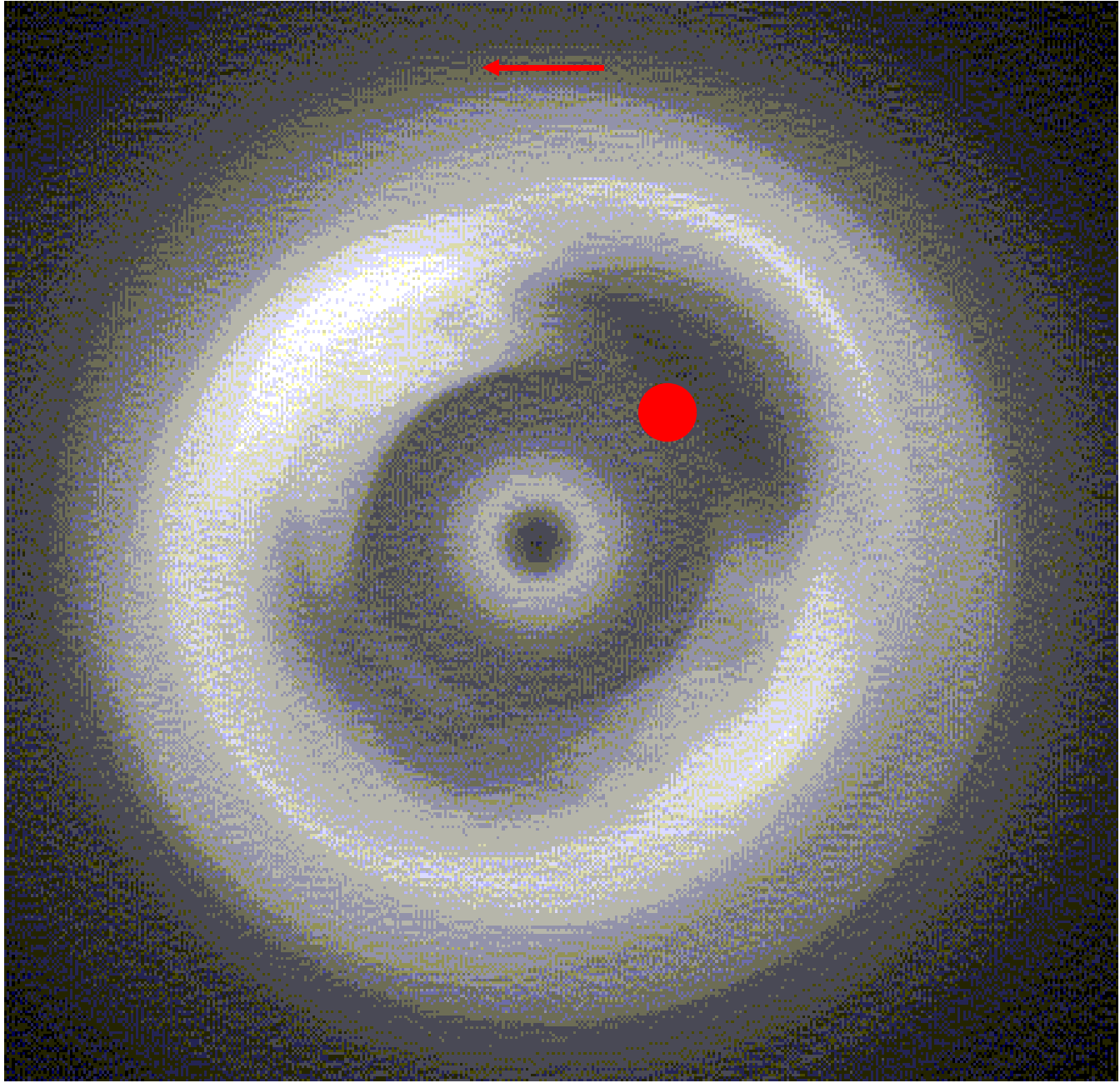

Fig.1. Simulated surface density of circumstellar dust disk with one Jupiter-like planet (=red circle). Orbital radius is 30 pixels (1 pixel = 2.5-3 AU for the Vega).

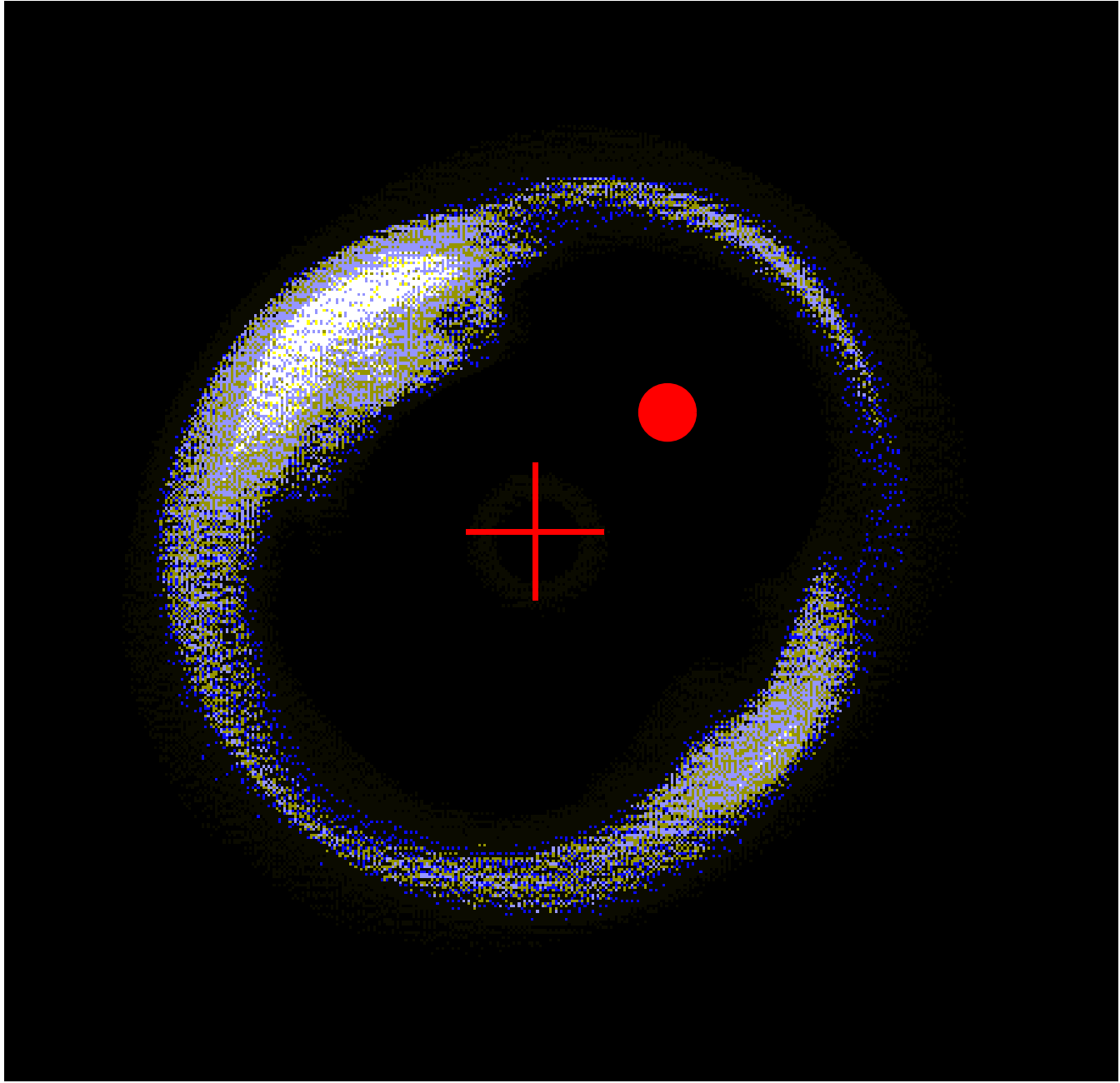

Fig.2. Most dense part of circumstellar dust disk from Fig.1. The numerical data for Figs.1-2 was calculated on 01/16/2000 (see details of simulation in Gorkavyi et.al. 2000a,b).

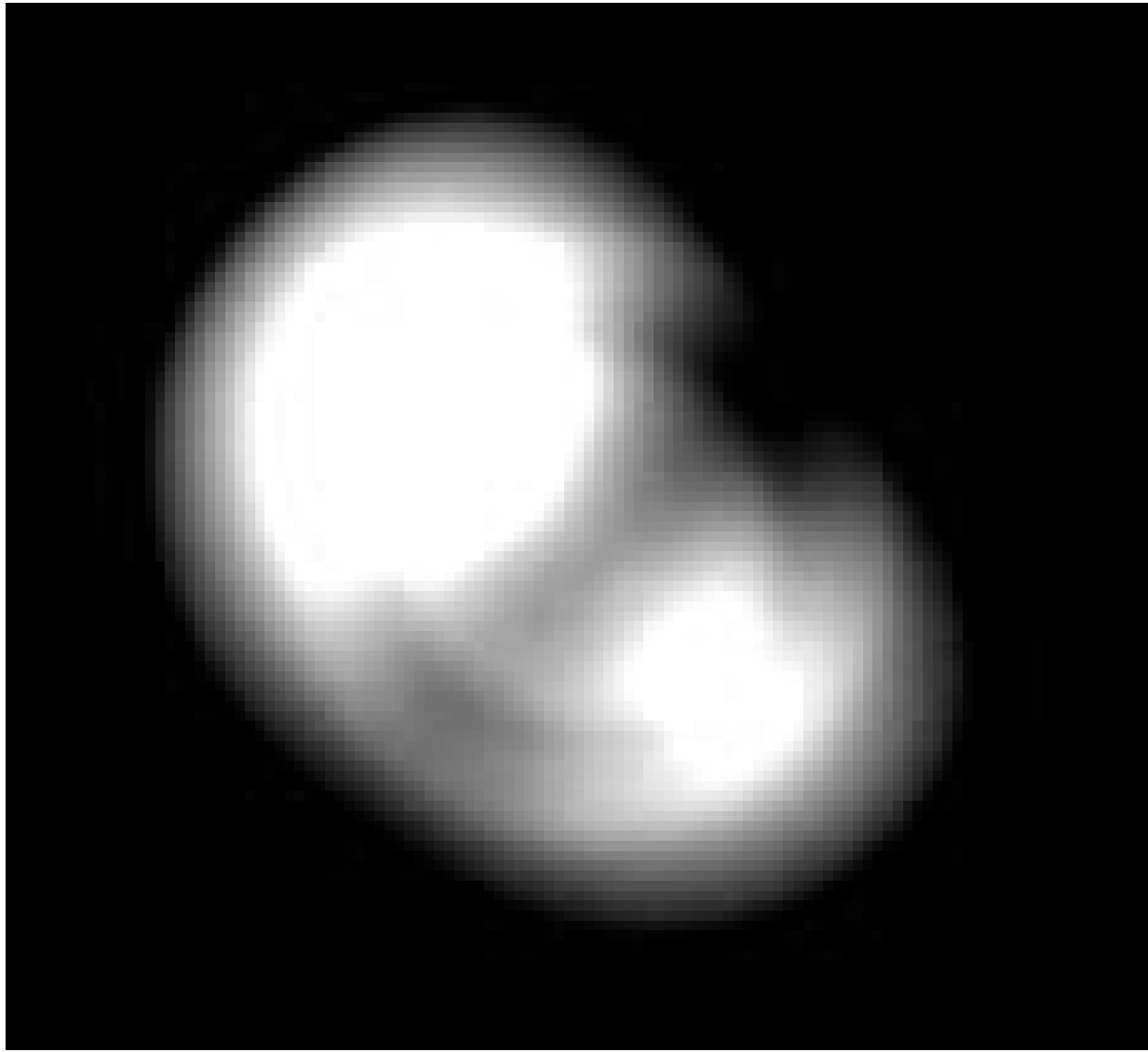

Fig.3. Thermal emission from circumstellar disk of the Vega (model from Fig.1-2). To compare our simulation with observation by Holland et al. 1998, we used averaging with an appropriate beam size (25 pixels). (Gorkavyi et.al. 2000a)

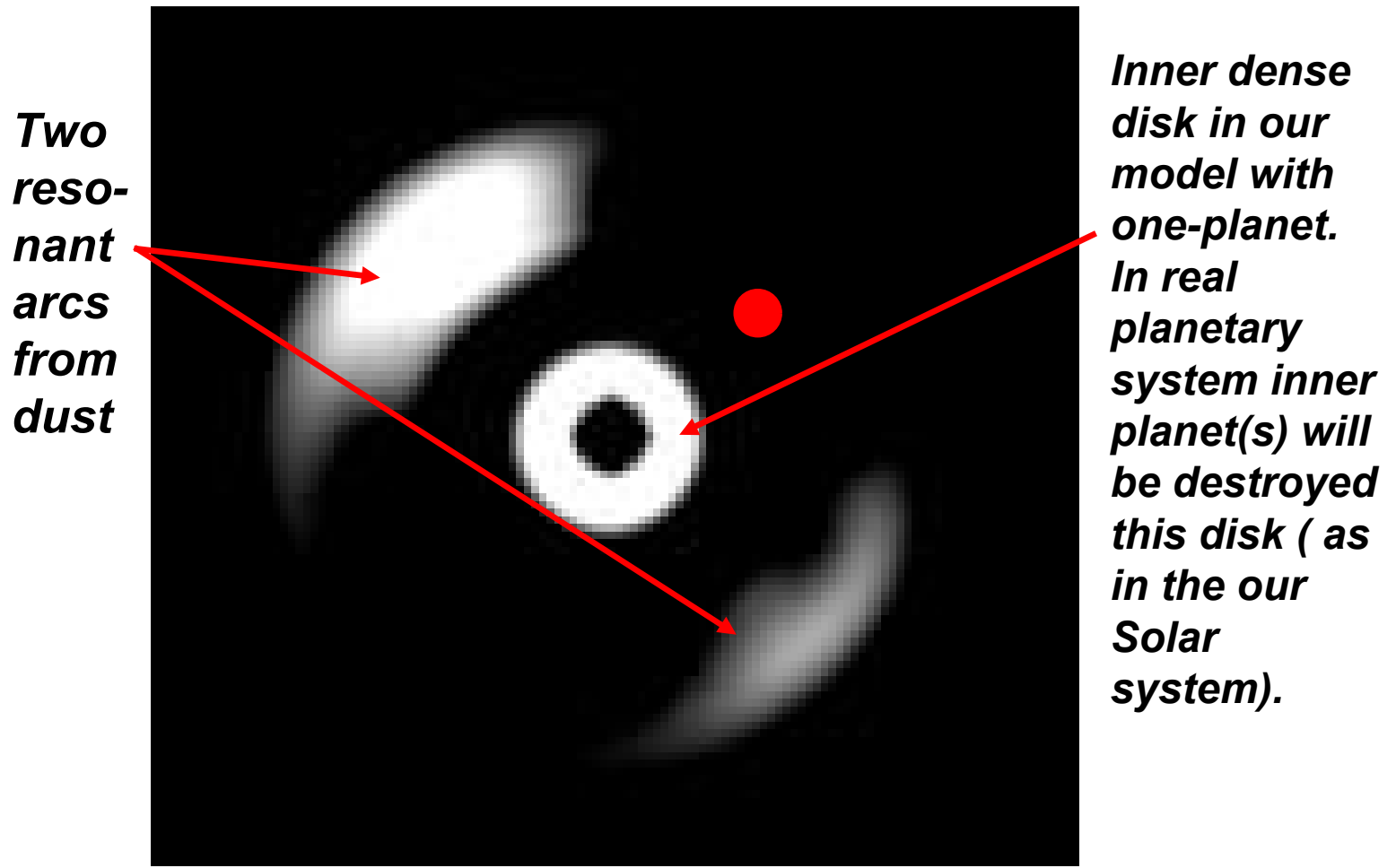

Fig.4. Thermal emission from circumstellar disk (model from Fig.1,2). 5 pixel' averaging was used for comparison of our simulation with observation by Koerner, Sargent and Ostroff, 2001.